\documentclass[12pt]{article}
\usepackage{times}
\usepackage{geometry}
\usepackage[utf8]{inputenc}
\usepackage{enumitem,amssymb}
\usepackage{ragged2e}
\usepackage{graphicx}
\usepackage{natbib}
\usepackage{aas_macros}
\newlist{thematic}{itemize}{8}
\setlist[thematic]{label=$\square$}
\usepackage{pifont}
\newcommand{\cmark}{\ding{51}}%
\newcommand{\done}{\rlap{$\square$}{\raisebox{2pt}{\large\hspace{1pt}\cmark}}%
\hspace{-2.5pt}}

\begin{document}
\bibliographystyle{aasjournal}
\raggedright
\huge
Astro2020 Science White Paper\linebreak

Envisioning the next decade of Galactic Center science: a laboratory for the study of the physics and astrophysics of supermassive black holes  \linebreak
\normalsize

\noindent \textbf{Thematic Areas:} \hspace*{60pt} $\square$ Planetary Systems \hspace*{10pt} $\square$ Star and Planet Formation \hspace*{20pt}\linebreak
$\done$ Formation and Evolution of Compact Objects \hspace*{31pt} $\done$ Cosmology and Fundamental Physics \linebreak
  $\square$  Stars and Stellar Evolution \hspace*{1pt} $\done$ Resolved Stellar Populations and their Environments \hspace*{40pt} \linebreak
  $\done$    Galaxy Evolution   \hspace*{45pt} $\square$             Multi-Messenger Astronomy and Astrophysics \hspace*{65pt} \linebreak
  
\textbf{Principal Author:}

Name: Tuan Do \& Andrea Ghez
 \linebreak						
Institution: UCLA 
 \linebreak
Email: tdo@astro.ucla.edu
 \linebreak
Phone:  
 \linebreak
 
\textbf{Co-authors:} (names and institutions)
Jessica R. Lu (UC Berkeley), Mark Morris (UCLA), Matthew Hosek Jr. (UCLA), Aurelien Hees (SYRTE, Observatoire de Paris, Universit\'e PSL, CNRS, Sorbonne Universit\'e), Smadar Naoz (UCLA), Anna Ciurlo (UCLA),  Philip J. Armitage	(Stony Brook University and Flatiron Institute), Rachael L Beaton (Princeton University), Eric Becklin (UCLA), Andrea Bellini (STScI), Rory O. Bentley (UCLA), Joss Bland-Hawthorn	(Sydney Institute for Astronomy, University of Sydney), Sukanya Chakrabarti (RIT), Zhuo Chen (UCLA), Devin S. Chu (UCLA), Arezu Dehghanfar (IPAG), Charles F. Gammie (University of Illinois at Urbana-Champaign), Abhimat K. Gautam (UCLA), Reinhard Genzel (Max Planck Institute for Extraterrestrial Physics), Jenny Greene (Princeton University), Daryl Haggard (McGill University), Joseph Hora (Center for Astrophysics, Harvard \& Smithsonian), Wolfgang E. Kerzendorf (New York University), Mattia Libralato (STScI),  Shogo Nishiyama (Miyagi University of Education), Kelly Kosmo O'Neil (UCLA), Feryal Ozel (University of Arizona), Thibaut Paumard (LESIA, Observatoire de Paris, Université PSL, CNRS, Sorbonne Université), Hagai B. Perets (Technion - Israel Institute of Technology), Dimitrios Psaltis (University of Arizona), Eliot Quataert (UC Berkeley), Enrico Ramirez-Ruiz (UCSC), R Michael Rich (UCLA), Fred Rasio (CIERA, Northwestern University), Shoko Sakai (UCLA),  Rainer Schoedel (Instituto de Astrofísica de Andalucía), Howard Smith (Center for Astrophysics, Harvard \& Smithsonian), Nevin N. Weinberg (MIT), Gunther Witzel (MPIfR, Bonn)
\linebreak
\justify
\textbf{Abstract  (optional):}
As the closest example of a galactic nucleus, the Galactic center (GC) presents an exquisite laboratory for learning about supermassive black holes (SMBH) and their environment. 
We describe several exciting new research directions that, over the next 10 years, hold the potential to answer some of the biggest scientific questions raised in recent decades:
Is General Relativity (GR) the correct description for supermassive black holes? What is the nature of star formation in extreme environments? How do stars and compact objects dynamically interact with the supermassive black hole? What physical processes drive gas accretion in low-luminosity black holes? 
We describe how the high sensitivity, angular resolution, and astrometric precision offered by the next generation of large ground-based telescopes with adaptive optics will help us answer these questions.
First, it will be possible to obtain precision measurements of stellar orbits in the Galaxy's central potential, providing both tests of GR in the unexplored regime near a SMBH and measurements of the extended dark matter distribution that is predicted to exist at the GC. The orbits of these stars will also allow us to measure the spin of the SMBH. 
Second, we will probe stellar populations at the GC to significantly lower masses than are possible today, down to the brown dwarf limit. Their structure and dynamics will provide an unprecedented view of the stellar cusp around the SMBH and will distinguish between models of star formation in the extreme environment of galactic nuclei. This increase in depth will also allow us to measure the currently unknown population of compact remnants at the GC by observing their effects on luminous sources. Third, uncertainties on the mass of and distance to the SMBH can be improved by a factor of $\sim$10. Finally, we can also study the near-infrared accretion onto the black hole at unprecedented sensitivity and time resolution, which can reveal the underlying physics of black hole accretion.

\begin{figure}[h!]
\begin{center}
\includegraphics[width=6.5in]{fig1.pdf}
\end{center}
\end{figure}

\pagebreak
The proximity of the Galactic center (GC) presents a unique opportunity to study a galactic nucleus at angular resolution orders of magnitude better than for any other galaxy. 
Over the last two decades, 10m class telescopes have enabled diffraction-limited imaging and spectroscopic observations of stars within the Milky Way's central cluster; our vision and  understanding of the center of the Milky Way have been fundamentally transformed. These include (1) the existence of a supermassive black hole (SMBH) at the center of our Galaxy \citep{Schodel:2002qq, Schodel:2003ek, Ghez:2003ul, Ghez:2005tx, Ghez:2008tg, Genzel:2010xq, Boehle:2016rt, Gillessen:2009uo, Gillessen:2017ai}, proving both that such exotic objects exist in the Universe, and that most galaxies harbor such objects at their centers; (2) the detection of the relativistic redshift of the star S0-2 during its closest approach to the black hole in 2018, providing the first test of General Relativity (GR) around a SMBH with stellar orbits \citep[][Do et al. submitted]{GRAVITY-Collaboration:2018cq}, and (3) stellar populations near the black hole whose nature appears to contradict our expectations for what is physically possible in such an environment, challenging our notions of how black holes and galaxies form and grow over time \citep[e.g.][]{Ghez:2003ul, Paumard:2006sh, Do:2009jk,Gallengo-cano18}. Additional progress now appears close to being stalled by the tremendous source confusion present within the angular resolution of our measurements.

\begin{figure}[h]
\begin{center}
\includegraphics[width=5in]{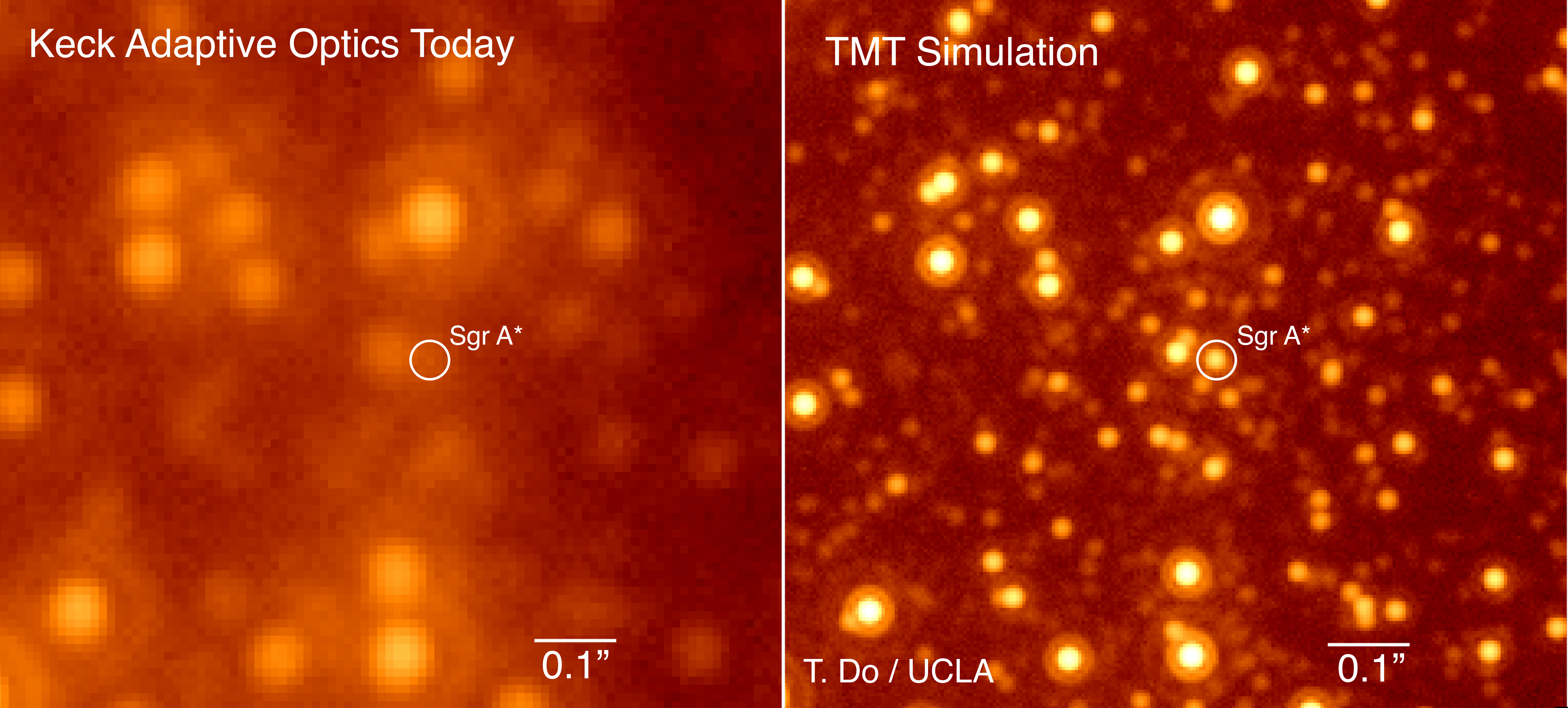}
\vspace{-0.3cm}
\caption{\footnotesize Simulation of the central 0.5$^{\prime\prime}$ (0.02 pc radius) around the supermassive black hole, Sgr A* (white circle), with current adaptive optics with the Keck Telescopes (left) compared to that with the Thirty Meter Telescope (right). The greater resolving power of ELTs, along with improved adaptive optics correction, will likely allow us to detect a factor of $>$10 times more stars in this region. Simulation details can be found in \citet{Do:2017lr}.}
\label{fig:TMT_improvements}
\end{center}
\end{figure}

The next generation of extremely large telescopes (ELTs) will carry this progress, with clearer vision,  a factor of three closer to SgrA*.
The greater flux sensitivity and spatial resolution of these facilities will allow for several orders of magnitude more stars to be monitored, including those located much closer to the SMBH, with greater accuracy in both position and velocity measurements (Fig. \ref{fig:TMT_improvements}). This will open new ways to use orbits to study the nature of gravity, as well as the effects of a SMBH on its environment. In this contribution, we highlight several important science questions that will help to drive GC research in the coming decade. 

\vspace{-0.9cm}

\noindent\underline{{\bf Is General Relativity the correct description of supermassive black holes?}}
While the theory of GR is well tested on Earth, in the solar system, and for certain astrophysical settings (e.g. binary pulsars) there are important regimes where GR has not been challenged at high precision, including that of supermassive black holes. 
It is unclear over what scale or regime GR might break down, so it is important to test it and other theories of gravity in new ways \citep[e.g.][]{Rubilar:2001zp, Weinberg:2005bf, Will:2008ud, Hees:2017yg}.  For example, two decades of monitoring the relativistic redshift of the short orbital period stars near the GC shows that the star S0-2/S2 is just now beginning to deviate from a Keplerian orbit \citep[][Do et al. submitted]{GRAVITY-Collaboration:2018cq}. 
In the coming decade, it will be possible to measure higher order changes to the orbit of S0-2/S2 and other stars. \textit{This would constitute one of the most important experiments in astrophysics, as such measurements will test both the Einstein equivalence principle and the form of the Kerr metric around a BH, allowing a search for a massive graviton or additional interactions (like scalar fields).}

With improved astrometric and radial velocity precision, it will be possible to detect GR ef-
fects that have no analog in classical dynamics such as the precession of the periapse \citep[e.g.][]{Do:2017lr} and the Lense-Thirring or frame dragging effect, which is due to the spin of the black hole \citep{Grould:2017zl}. The potential to monitor orbits of more stars is also key to identify GR effects. With precise measurements of multiple stars, it should not only be possible to separate out relativistic effects from noise, but also from the extended mass distribution (which is highly correlated if only one star is used).

\begin{figure}[th]
\begin{center}
\vspace{-0.4cm}
\includegraphics[width=6.5in]{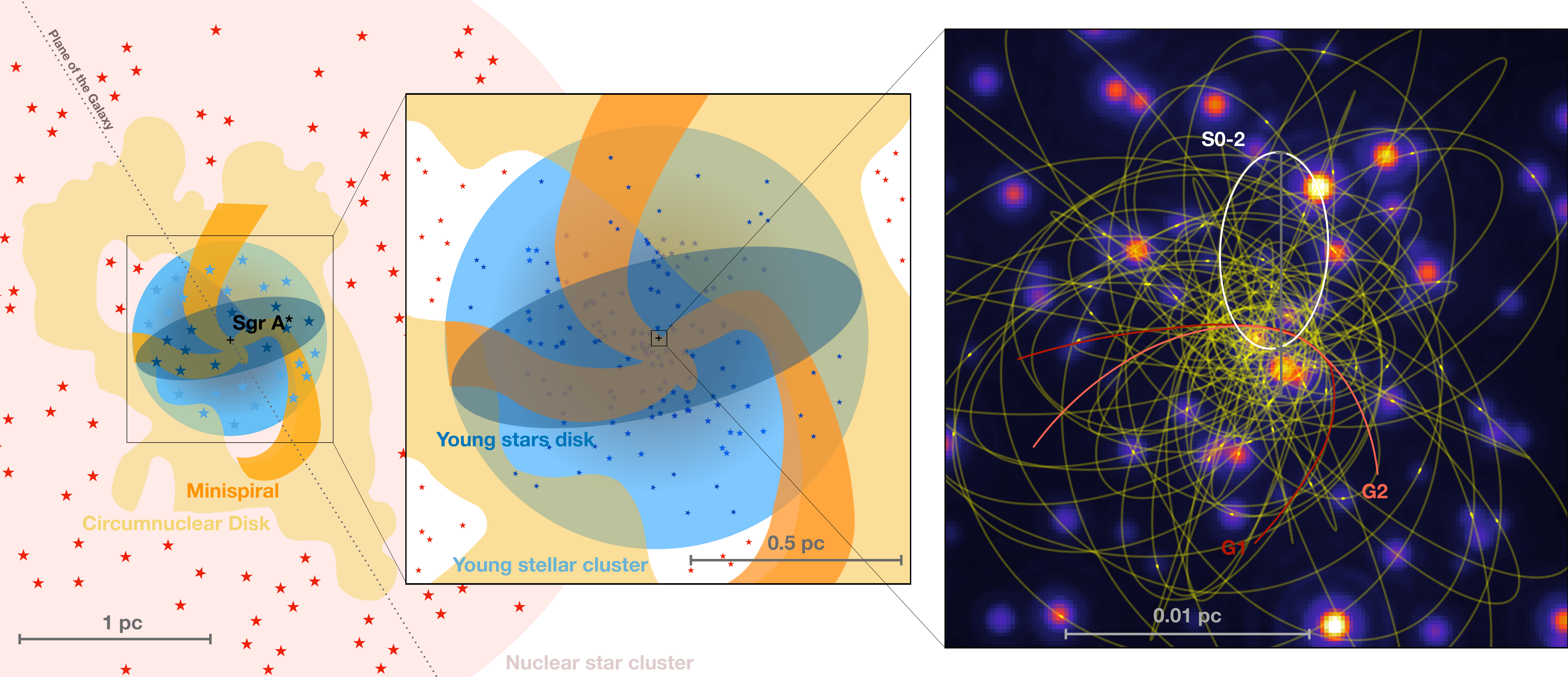}
\vspace{-0.9cm}
\caption{\footnotesize The GC is a complex environment, containing the densest star cluster known in the Milky Way with multiple stellar populations as well as various gas structures. These diagrams show a zoom-in from \textbf{(left)}: parsec scales, which is dominated by old stars (red, 1-10 Gyr old) and gas, to \textbf{(middle)}: 0.5-pc scale, which hosts young stars (blue, 4-6 Myr), to \textbf{(right)}: 0.05-pc scale where the stellar orbits can be used to test GR (simulated). Today, only one star (white, S0-2/S2) has the astrometric and radial velocity precision to constrain GR. ELTs will discover stars with orbital periods as small as 1-2 years given their increased sensitivity and angular resolution. With ELTs, it should be possible to use $\sim100$ stars jointly for GR tests. This region is also ideal for studying the interaction of stars and compact objects with the supermassive black hole. G1 and G2 are two sources (orange) known today to show tidal interactions. Image credit: A. Ciurlo \& \citet{Do:2017lr}.}
\vspace{-0.95cm}
\label{fig:orbits}
\end{center}
\end{figure}

GR effects (or those of other theories of gravity) are stronger for stars with a short period and high eccentricity; thus, detecting these sources is crucial to effective tests of gravity. Currently, the shortest-period star known today among the inner-arcsecond sources surrounding SgrA* has a period of 11.5 years \citep[S0-102,][]{Meyer12}. 
Based on our knowledge of the stellar luminosity function, we predict that by going 5 magnitudes fainter than our confusion limit today, there will likely be multiple sources with periods as short as 1-2 years; these stars will be instrumental for testing GR (Fig. \ref{fig:orbits}).

\noindent\underline{{\bf What is the distribution of dark mass at the GC?}}
Observations of deviations from Keplerian orbits will also allow the detection of the extended mass around the SMBH. The extended dark mass around the SMBH can consist of both dark matter and compact objects such as neutron stars and stellar mass black holes.  

Dynamical friction is thought to concentrate compact objects near the black hole in a dense dark cusp, where sources may eventually merge with the black hole producing gravitational waves \citep{Alexander17}.  
It is expected that the dark matter density profile in galactic nuclei should exhibit a sharp spike near the central SMBH \citep[e.g.][]{Gondolo:1999kk}, though this has never been observationally demonstrated.
To date, only upper limits on the extended mass distribution have been achieved, and only the largest of such possible density spikes have been ruled out \citep{Lacroix:2018lh}.
ELTs will be able to detect the influence of the extended mass distribution on stellar orbits as long as the dark mass within 0.01 pc is at least 2500 M$_{\odot}$ \citep[an order of magnitude more sensitive than current constraints;][]{Boehle:2016rt, Gillessen:2017ai}.

In addition to using the stellar orbits to measure extended mass, there are other ways to detect compact objects. For example: (1) gravitational lensing from stellar mass black holes can be used to determine their masses and number density; (2) if compact objects are in binary systems with luminous companions, they will be detectable via astrometric and radial velocity variations; (3) the binary fraction and period distribution of stars can also be used as an indirect means for constraining the existence of a dark cusp of compact objects. Over time, these sources should break up wide-binary systems via two-body interactions, and will have predictable effects on the number and separation of typical binaries that can survive as long as the age of the cluster \citep[e.g.][]{Alexander17,Pfuhl14}. These methods are only beginning to be employed today, and their effectiveness scales with the number of sources that can be monitored as well as the time baseline of observations.

\begin{figure}[hbt]
\center
\vspace{-0.5cm}
\includegraphics[width=6.75in]{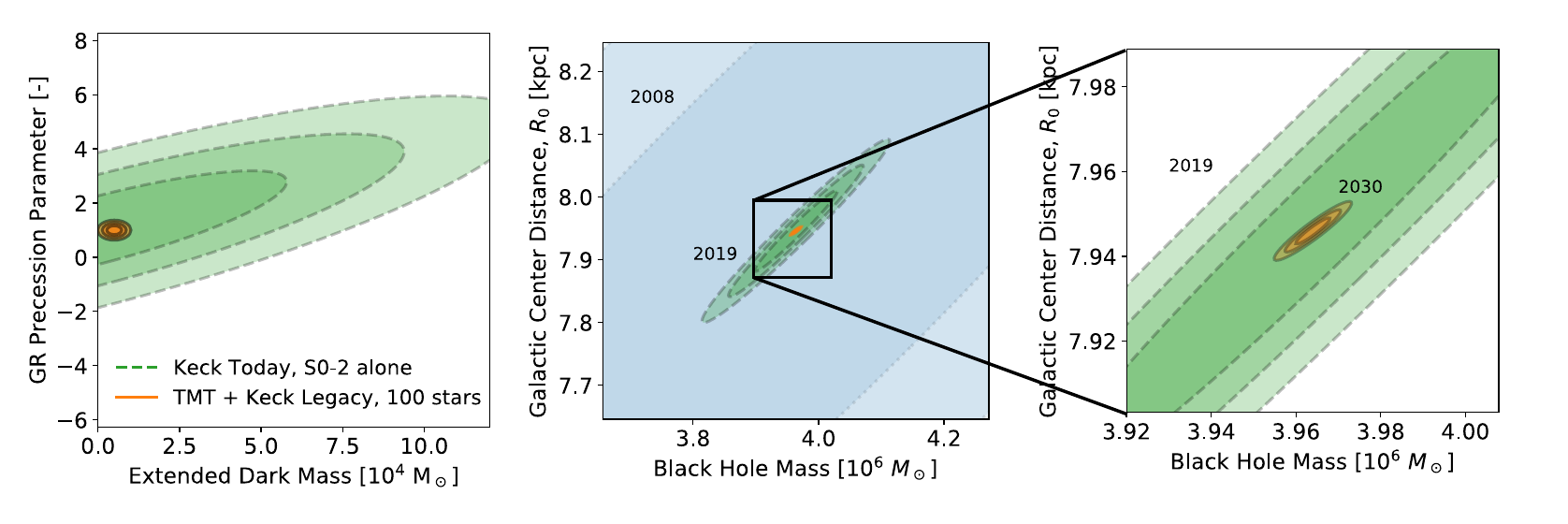}
\vspace{-0.8cm}
\caption{\footnotesize\textbf{Left}: Joint probability distribution function for the GR precession signal (1 for GR, 0 for Newtonian) and the total extended mass within $\sim100$ AU of the black hole (with 1,2,3 $\sigma$ contours) for Keck today (green) and simulations with TMT (orange). With only one star, these two parameters are highly correlated. The addition of other stars are required to break this degeneracy and obtain precise measurements of both these parameters. \textbf{Center \& Right:} Joint probability distribution of black hole mass and distance to the GC. In the past 10 years, the uncertainties in these parameters have improved by a factor of 10 (2008: blue, 2019: green). With three years of observations from ELTs of $\sim100$ stars and legacy data from telescopes today, another factor of $>10$ improvement in the uncertainties is possible (orange). See \citet{Do:2017lr} for details.}
\vspace{-0.6cm}
\label{fig:gr_extm}
\end{figure}

\noindent\underline{{\bf Improved Black Hole Mass and Galactic Center Distance Measurements}}
The measurement of stellar orbits around Sgr A* is the most precise method for measuring the distance to the GC. Current uncertainties are below 0.1\% using the star S0-2/S2 \citep[][Do et al. submitted]{GRAVITY-Collaboration:2018cq}. This measurement can be improved by an order of magnitude ($\sigma_{R_o} < 5$ pc) by including more stars and higher precision radial velocity and astrometric measurements (Fig. \ref{fig:gr_extm}). These improvements should remove the contribution from the uncertainty of the GC distance from calculations involving the structure or scale of the Milky Way \citep[e.g.,][]{Olling:2000rp}.

\noindent\underline{{\bf What is the nature of star formation in extreme environments?}}
A long standing mystery at the GC is the paradox of youth: there is a young star cluster (4-6 Myr) concentrated within just 0.5 pc of the supermassive black hole \citep[e.g.][]{Stostad15}. In fact, some of the closest stars orbiting the black hole are young B-type stars. Star formation should be inhibited at the GC as the tidal forces from the black hole should prevent molecular clouds from collapsing to form stars. 

\begin{figure}[h]
\begin{center}
\vspace{-0.5cm}
\includegraphics[scale=0.25]{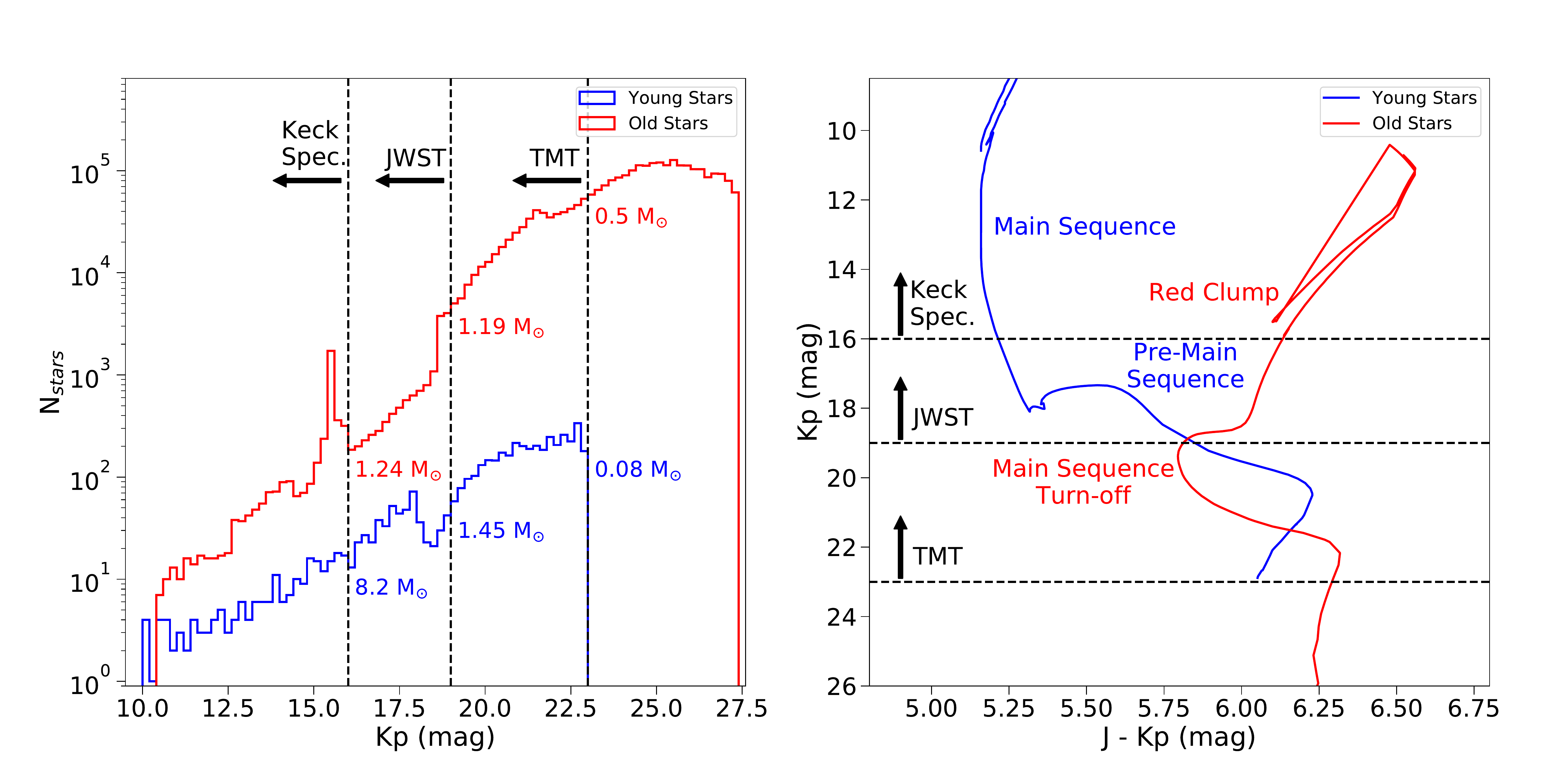}
\vspace{-0.5cm}
\caption{\footnotesize{\bf Left:} Luminosity function of the young (blue, 4-6 Myr) and old (red, 1-10 Gyr) stellar population at the GC. Current spectroscopy from 8-10 m telescopes are limited only to the most massive young stars or red giants. ELTs will provide many magnitudes of improvement in spectroscopic sensitivity, which will reach down to the hydrogen burning limit for the young stars (i.e. brown dwarfs). {\bf Right:} Isochrones of the young and old stars, showing important regions such as the pre-main sequence turn-on and the main sequence turn-off. It is crucial to measure these stages of star formation to understand the nature of star formation at the GC.}
\vspace{-1cm}
\label{fig:pops}
\end{center}
\end{figure}

Studies of star formation at the GC require measurements of fundamental parameters such as the initial mass function, binary fraction, spatial distribution, and dynamical properties in order to provide constraints on star formation models. Today, the estimates of these parameters are driven by the population of young stars that can be observed, which are the most massive stars ($8-10 M_\odot$). Based on this population of stars, we find some unusual properties such as a top heavy mass function and about 20\% of these stars are orbiting the black hole in a stellar disk \citep{Paumard:2006sh,Lu:2009rq}. These observations support the hypothesis that the stars formed in-situ via fragmentation of a massive accretion disk around the black hole 4-6 Myr ago \citep[e.g.][]{Nayakshin07,Bonnell08}. 
However, many questions remain: what is the physics that govern the accretion disk and what are the details of star formation in such an environment? Why are 50-80\% of the young stars not members of the stellar disk \citep{Bartko09,Yelda14}?  Answers to these questions require observations that are out of reach today. Observations such as the turn-over mass of the mass function or the binary fraction are inaccessible today because of lack of sensitivity and angular resolution. For example, Keck spectroscopy is currently limited by the sensitivity, with magnitude limits of Kp $<$ 16 for several hours of integration time \citep[primarily young OB stars with mass $> 10$ $M_\odot$ and old Red Clump stars][]{Do:2013il}. In contrast, ELTs such as TMT (with the IRIS instrument) are predicted to achieve high SNR spectroscopy for Kp = 22 - 23 mag stars (Fig. \ref{fig:pops}), corresponding to masses of 0.5 M$_{\odot}$ for the old star population and 0.08 M$_{\odot}$ for the young star population.  With imaging we will observe young brown dwarfs at K = 24-26 mag. Such observations will allow for the deepest measurement of the stellar IMF in the extreme GC environment. Measuring where the peak in the GC IMF occurs relative to local star forming regions yields critical insight into the physics driving star formation in the region \cite[e.g.][]{Offner:2014vn, Krumholz:2014ne}.

\noindent\underline{\textbf{How do stars and compact objects dynamically interact with the supermassive black hole?}} \\
The interaction between supermassive black holes and dense stellar clusters is thought to drive many important dynamical processes such as the tidal disruption of stars and increasing the merger rates of compact objects that produce gravitational waves. These phenomena can only be observed in detail at the GC. Currently, groups are exploring the effects of binaries (both stellar and compact objects) in shaping the physical properties of the stellar distribution and creation of merger products. For example, binaries are invoked to explain some long-standing observational puzzles such as hypervelocity stars, the young stars in the S-cluster, the dark cusp, etc., \citep[e.g.,][]{Hills88,Yu+03,Naoz+18,Stephan+19}. Furthermore, it has been suggested that compact object binaries in the GC are a potential source of gravitational wave emission \citep[e.g.][]{OLeary+09,Antonini12,Hoang+19}. Observations of tidally interacting objects such as G1 and G2 \citep[e.g.][]{Gillessen12,Witzel17} show that we can potentially follow their evolution in detail. 

\noindent\underline{{\bf What physical processes drive gas accretion in low-luminosity black holes?}}
The accretion flow onto Sgr A* is observable at radio to X-ray wavelengths as a very red source that varies continuously on time scales from minutes to hours \citep[e.g.][]{Do09,Dodds-Eden11,Witzel18}.  The latter time scale corresponds to the delay above which the variations become uncorrelated. Since many physical processes having stochastic behavior give rise to a red-noise power spectrum, the variability has not yet led to a conclusive understanding of the underlying cause of the variations.  Ideas include magnetic reconnection events, inhomogeneous mass accretion, and disk instabilities -- all perhaps modulated by gravitational lensing.  However, the shortest time scales can elucidate and delimit the physical scales of the key interactions.  For example, 8.5 minutes corresponds to the orbital period of the innermost stable circular orbit around a black hole having the mass of Sgr A*, and a spin rate of 92\% of the maximum possible value, so variability properties at the shortest time scales can certainly provide interesting constraints on the dynamics of the varying regions as well as on the rise times of the variable "events".  With the greatly enhanced sensitivity of the ELTs, it will be possible to monitor the near-infrared emission from Sgr A* with a cadence below 1 second ($>$ 2 orders of magnitude improvement in time resolution), allowing probing variations on physical scales well below any other experiment.  This capability would open a new window to the underlying physics of the variable accretion flow.

\noindent\underline{\textbf{Recommendations}}
\noindent\textbf{-} The science cases presented above require high angular resolution to provide the sensitivity and precision to make accurate measurements of stellar orbits in the most crowded region of the Milky Way. long term, sustained, astrometric precision of $<$ 30 micro-arcseconds per epoch of observation will be required to measure the subtle effects of GR on the orbits. Improvement in spectroscopic sensitivity to K = 22-23 mag will be necessary to discover stars with several-year orbital periods around the black hole for GR test and to reveal the nature of star formation in this environment. These requirements can be met by the upcoming ELTs. \\
\noindent\textbf{-} ELTs with adaptive optics on both the northern and southern hemispheres will be ideal for the robustness and efficiency of the experiments outlined above. Separation in longitude will allow for long-time baseline observations that will be important for Sgr A* variability studies. \\
\noindent\textbf{-} As science programs become larger (such those proposed here), it becomes important to sustain large collaborations and datasets. The current NSF funding model has no research centers in the Astronomy Division. Such centers will become important in the era of ELTs.

\pagebreak
\bibliography{GC}

\begin{thebibliography}{}
\expandafter\ifx\csname natexlab\endcsname\relax\def\natexlab#1{#1}\fi

\bibitem[{{Alexander}(2017)}]{Alexander17}
{Alexander}, T. 2017, \araa, 55, 17

\bibitem[{{Antonini} \& {Perets}(2012)}]{Antonini12}
{Antonini}, F., \& {Perets}, H.~B. 2012, \apj, 757, 27

\bibitem[{{Bartko} {et~al.}(2009){Bartko}, {Martins}, {Fritz}, {Genzel},
  {Levin}, {Perets}, {Paumard}, {Nayakshin}, {Gerhard}, {Alexander},
  {Dodds-Eden}, {Eisenhauer}, {Gillessen}, {Mascetti}, {Ott}, {Perrin},
  {Pfuhl}, {Reid}, {Rouan}, {Sternberg}, \& {Trippe}}]{Bartko09}
{Bartko}, H., {Martins}, F., {Fritz}, T.~K., {et~al.} 2009, \apj, 697, 1741

\bibitem[{{Boehle} {et~al.}(2016){Boehle}, {Ghez}, {Sch{\"o}del}, {Meyer},
  {Yelda}, {Albers}, {Martinez}, {Becklin}, {Do}, {Lu}, {Matthews}, {Morris},
  {Sitarski}, \& {Witzel}}]{Boehle:2016rt}
{Boehle}, A., {Ghez}, A.~M., {Sch{\"o}del}, R., {et~al.} 2016, \apj, 830, 17

\bibitem[{{Bonnell} \& {Rice}(2008)}]{Bonnell08}
{Bonnell}, I.~A., \& {Rice}, W.~K.~M. 2008, Science, 321, 1060

\bibitem[{{Do} {et~al.}(2009{\natexlab{a}}){Do}, {Ghez}, {Morris}, {Lu},
  {Matthews}, {Yelda}, \& {Larkin}}]{Do:2009jk}
{Do}, T., {Ghez}, A.~M., {Morris}, M.~R., {et~al.} 2009{\natexlab{a}}, \apj,
  703, 1323

\bibitem[{{Do} {et~al.}(2009{\natexlab{b}}){Do}, {Ghez}, {Morris}, {Yelda},
  {Meyer}, {Lu}, {Hornstein}, \& {Matthews}}]{Do09}
---. 2009{\natexlab{b}}, \apj, 691, 1021

\bibitem[{{Do} {et~al.}(2017){Do}, {Hees}, {Dehghanfar}, {Ghez}, \&
  {Wright}}]{Do:2017lr}
{Do}, T., {Hees}, A., {Dehghanfar}, A., {Ghez}, A., \& {Wright}, S. 2017, ArXiv
  e-prints, arXiv:1711.06389

\bibitem[{{Do} {et~al.}(2013){Do}, {Lu}, {Ghez}, {Morris}, {Yelda}, {Martinez},
  {Wright}, \& {Matthews}}]{Do:2013il}
{Do}, T., {Lu}, J.~R., {Ghez}, A.~M., {et~al.} 2013, \apj, 764, 154

\bibitem[{{Dodds-Eden} {et~al.}(2011){Dodds-Eden}, {Gillessen}, {Fritz},
  {Eisenhauer}, {Trippe}, {Genzel}, {Ott}, {Bartko}, {Pfuhl}, {Bower},
  {Goldwurm}, {Porquet}, {Trap}, \& {Yusef-Zadeh}}]{Dodds-Eden11}
{Dodds-Eden}, K., {Gillessen}, S., {Fritz}, T.~K., {et~al.} 2011, \apj, 728, 37

\bibitem[{{Gallego-Cano} {et~al.}(2018){Gallego-Cano}, {Sch{\"o}del}, {Dong},
  {Nogueras-Lara}, {Gallego-Calvente}, {Amaro-Seoane}, \&
  {Baumgardt}}]{Gallengo-cano18}
{Gallego-Cano}, E., {Sch{\"o}del}, R., {Dong}, H., {et~al.} 2018, \aap, 609,
  A26

\bibitem[{{Genzel} {et~al.}(2010){Genzel}, {Eisenhauer}, \&
  {Gillessen}}]{Genzel:2010xq}
{Genzel}, R., {Eisenhauer}, F., \& {Gillessen}, S. 2010, Reviews of Modern
  Physics, 82, 3121

\bibitem[{{Ghez} {et~al.}(2005){Ghez}, {Salim}, {Hornstein}, {Tanner}, {Lu},
  {Morris}, {Becklin}, \& {Duch{\^e}ne}}]{Ghez:2005tx}
{Ghez}, A.~M., {Salim}, S., {Hornstein}, S.~D., {et~al.} 2005, \apj, 620, 744

\bibitem[{{Ghez} {et~al.}(2003){Ghez}, {Duch{\^e}ne}, {Matthews}, {Hornstein},
  {Tanner}, {Larkin}, {Morris}, {Becklin}, {Salim}, {Kremenek}, {Thompson},
  {Soifer}, {Neugebauer}, \& {McLean}}]{Ghez:2003ul}
{Ghez}, A.~M., {Duch{\^e}ne}, G., {Matthews}, K., {et~al.} 2003, \apjl, 586,
  L127

\bibitem[{{Ghez} {et~al.}(2008){Ghez}, {Salim}, {Weinberg}, {Lu}, {Do}, {Dunn},
  {Matthews}, {Morris}, {Yelda}, {Becklin}, {Kremenek}, {Milosavljevic}, \&
  {Naiman}}]{Ghez:2008tg}
{Ghez}, A.~M., {Salim}, S., {Weinberg}, N.~N., {et~al.} 2008, \apj, 689, 1044

\bibitem[{{Gillessen} {et~al.}(2009){Gillessen}, {Eisenhauer}, {Fritz},
  {Bartko}, {Dodds-Eden}, {Pfuhl}, {Ott}, \& {Genzel}}]{Gillessen:2009uo}
{Gillessen}, S., {Eisenhauer}, F., {Fritz}, T.~K., {et~al.} 2009, \apjl, 707,
  L114

\bibitem[{{Gillessen} {et~al.}(2012){Gillessen}, {Genzel}, {Fritz}, {Quataert},
  {Alig}, {Burkert}, {Cuadra}, {Eisenhauer}, {Pfuhl}, {Dodds-Eden}, {Gammie},
  \& {Ott}}]{Gillessen12}
{Gillessen}, S., {Genzel}, R., {Fritz}, T.~K., {et~al.} 2012, \nat, 481, 51

\bibitem[{{Gillessen} {et~al.}(2017){Gillessen}, {Plewa}, {Eisenhauer}, {Sari},
  {Waisberg}, {Habibi}, {Pfuhl}, {George}, {Dexter}, {von Fellenberg}, {Ott},
  \& {Genzel}}]{Gillessen:2017ai}
{Gillessen}, S., {Plewa}, P.~M., {Eisenhauer}, F., {et~al.} 2017, \apj, 837, 30

\bibitem[{{Gondolo} \& {Silk}(1999)}]{Gondolo:1999kk}
{Gondolo}, P., \& {Silk}, J. 1999, Physical Review Letters, 83, 1719

\bibitem[{{GRAVITY Collaboration} {et~al.}(2018){GRAVITY Collaboration},
  {Abuter}, {Amorim}, {Anugu}, {Baub{\"o}ck}, {Benisty}, {Berger}, {Blind},
  {Bonnet}, {Brandner}, {Buron}, {Collin}, {Chapron}, {Cl{\'e}net}, {Coud{\'e}
  du Foresto}, {de Zeeuw}, {Deen}, {Delplancke-Str{\"o}bele}, {Dembet},
  {Dexter}, {Duvert}, {Eckart}, {Eisenhauer}, {Finger}, {F{\"o}rster
  Schreiber}, {F{\'e}dou}, {Garcia}, {Garcia Lopez}, {Gao}, {Gendron},
  {Genzel}, {Gillessen}, {Gordo}, {Habibi}, {Haubois}, {Haug}, {Hau{\ss}mann},
  {Henning}, {Hippler}, {Horrobin}, {Hubert}, {Hubin}, {Jimenez Rosales},
  {Jochum}, {Jocou}, {Kaufer}, {Kellner}, {Kendrew}, {Kervella}, {Kok},
  {Kulas}, {Lacour}, {Lapeyr{\`e}re}, {Lazareff}, {Le Bouquin}, {L{\'e}na},
  {Lippa}, {Lenzen}, {M{\'e}rand}, {M{\"u}ller}, {Neumann}, {Ott}, {Palanca},
  {Paumard}, {Pasquini}, {Perraut}, {Perrin}, {Pfuhl}, {Plewa}, {Rabien},
  {Ram{\'{\i}}rez}, {Ramos}, {Rau}, {Rodr{\'{\i}}guez-Coira}, {Rohloff},
  {Rousset}, {Sanchez-Bermudez}, {Scheithauer}, {Sch{\"o}ller}, {Schuler},
  {Spyromilio}, {Straub}, {Straubmeier}, {Sturm}, {Tacconi}, {Tristram},
  {Vincent}, {von Fellenberg}, {Wank}, {Waisberg}, {Widmann}, {Wieprecht},
  {Wiest}, {Wiezorrek}, {Woillez}, {Yazici}, {Ziegler}, \&
  {Zins}}]{GRAVITY-Collaboration:2018cq}
{GRAVITY Collaboration}, {Abuter}, R., {Amorim}, A., {et~al.} 2018, ArXiv
  e-prints, arXiv:1807.09409

\bibitem[{{Grould} {et~al.}(2017){Grould}, {Vincent}, {Paumard}, \&
  {Perrin}}]{Grould:2017zl}
{Grould}, M., {Vincent}, F.~H., {Paumard}, T., \& {Perrin}, G. 2017, \aap, 608,
  A60

\bibitem[{{Hees} {et~al.}(2017){Hees}, {Do}, {Ghez}, {Martinez}, {Naoz},
  {Becklin}, {Boehle}, {Chappell}, {Chu}, {Dehghanfar}, {Kosmo}, {Lu},
  {Matthews}, {Morris}, {Sakai}, {Sch{\"o}del}, \& {Witzel}}]{Hees:2017yg}
{Hees}, A., {Do}, T., {Ghez}, A.~M., {et~al.} 2017, Physical Review Letters,
  118, 211101

\bibitem[{{Hills}(1988)}]{Hills88}
{Hills}, J.~G. 1988, \nat, 331, 687

\bibitem[{{Hoang} {et~al.}(2019){Hoang}, {Naoz}, {Kocsis}, {Farr}, \&
  {McIver}}]{Hoang+19}
{Hoang}, B.-M., {Naoz}, S., {Kocsis}, B., {Farr}, W., \& {McIver}, J. 2019,
  arXiv e-prints, arXiv:1903.00134

\bibitem[{{Krumholz}(2014)}]{Krumholz:2014ne}
{Krumholz}, M.~R. 2014, \physrep, 539, 49

\bibitem[{{Lacroix}(2018)}]{Lacroix:2018lh}
{Lacroix}, T. 2018, ArXiv e-prints, arXiv:1801.01308

\bibitem[{{Lu} {et~al.}(2009){Lu}, {Ghez}, {Hornstein}, {Morris}, {Becklin}, \&
  {Matthews}}]{Lu:2009rq}
{Lu}, J.~R., {Ghez}, A.~M., {Hornstein}, S.~D., {et~al.} 2009, \apj, 690, 1463

\bibitem[{{Meyer} {et~al.}(2012){Meyer}, {Ghez}, {Sch{\"o}del}, {Yelda},
  {Boehle}, {Lu}, {Do}, {Morris}, {Becklin}, \& {Matthews}}]{Meyer12}
{Meyer}, L., {Ghez}, A.~M., {Sch{\"o}del}, R., {et~al.} 2012, Science, 338, 84

\bibitem[{{Naoz} {et~al.}(2018){Naoz}, {Ghez}, {Hees}, {Do}, {Witzel}, \&
  {Lu}}]{Naoz+18}
{Naoz}, S., {Ghez}, A.~M., {Hees}, A., {et~al.} 2018, \apjl, 853, L24

\bibitem[{{Nayakshin} {et~al.}(2007){Nayakshin}, {Cuadra}, \&
  {Springel}}]{Nayakshin07}
{Nayakshin}, S., {Cuadra}, J., \& {Springel}, V. 2007, \mnras, 379, 21

\bibitem[{{Offner} {et~al.}(2014){Offner}, {Clark}, {Hennebelle}, {Bastian},
  {Bate}, {Hopkins}, {Moraux}, \& {Whitworth}}]{Offner:2014vn}
{Offner}, S.~S.~R., {Clark}, P.~C., {Hennebelle}, P., {et~al.} 2014, Protostars
  and Planets VI, 53

\bibitem[{{O'Leary} {et~al.}(2009){O'Leary}, {Kocsis}, \& {Loeb}}]{OLeary+09}
{O'Leary}, R.~M., {Kocsis}, B., \& {Loeb}, A. 2009, \mnras, 395, 2127

\bibitem[{{Olling} \& {Merrifield}(2000)}]{Olling:2000rp}
{Olling}, R.~P., \& {Merrifield}, M.~R. 2000, \mnras, 311, 361

\bibitem[{{Paumard} {et~al.}(2006){Paumard}, {Genzel}, {Martins}, {Nayakshin},
  {Beloborodov}, {Levin}, {Trippe}, {Eisenhauer}, {Ott}, {Gillessen}, {Abuter},
  {Cuadra}, {Alexander}, \& {Sternberg}}]{Paumard:2006sh}
{Paumard}, T., {Genzel}, R., {Martins}, F., {et~al.} 2006, \apj, 643, 1011

\bibitem[{{Pfuhl} {et~al.}(2014){Pfuhl}, {Alexander}, {Gillessen}, {Martins},
  {Genzel}, {Eisenhauer}, {Fritz}, \& {Ott}}]{Pfuhl14}
{Pfuhl}, O., {Alexander}, T., {Gillessen}, S., {et~al.} 2014, \apj, 782, 101

\bibitem[{{Rubilar} \& {Eckart}(2001)}]{Rubilar:2001zp}
{Rubilar}, G.~F., \& {Eckart}, A. 2001, \aap, 374, 95

\bibitem[{{Sch{\"o}del} {et~al.}(2003){Sch{\"o}del}, {Ott}, {Genzel}, {Eckart},
  {Mouawad}, \& {Alexander}}]{Schodel:2003ek}
{Sch{\"o}del}, R., {Ott}, T., {Genzel}, R., {et~al.} 2003, \apj, 596, 1015

\bibitem[{{Sch{\"o}del} {et~al.}(2002){Sch{\"o}del}, {Ott}, {Genzel},
  {Hofmann}, {Lehnert}, {Eckart}, {Mouawad}, {Alexander}, {Reid}, {Lenzen},
  {Hartung}, {Lacombe}, {Rouan}, {Gendron}, {Rousset}, {Lagrange}, {Brandner},
  {Ageorges}, {Lidman}, {Moorwood}, {Spyromilio}, {Hubin}, \&
  {Menten}}]{Schodel:2002qq}
---. 2002, \nat, 419, 694

\bibitem[{{Stephan} {et~al.}(2019){Stephan}, {Naoz}, {Ghez}, {Morris},
  {Ciurlo}, {Do}, {Breivik}, {Coughlin}, \& {Rodriguez}}]{Stephan+19}
{Stephan}, A.~P., {Naoz}, S., {Ghez}, A.~M., {et~al.} 2019, arXiv e-prints,
  arXiv:1903.00010

\bibitem[{{St{\o}stad} {et~al.}(2015){St{\o}stad}, {Do}, {Murray}, {Lu},
  {Yelda}, \& {Ghez}}]{Stostad15}
{St{\o}stad}, M., {Do}, T., {Murray}, N., {et~al.} 2015, \apj, 808, 106

\bibitem[{{Weinberg} {et~al.}(2005){Weinberg}, {Milosavljevi{\'c}}, \&
  {Ghez}}]{Weinberg:2005bf}
{Weinberg}, N.~N., {Milosavljevi{\'c}}, M., \& {Ghez}, A.~M. 2005, \apj, 622,
  878

\bibitem[{{Will}(2008)}]{Will:2008ud}
{Will}, C.~M. 2008, \apjl, 674, L25

\bibitem[{{Witzel} {et~al.}(2017){Witzel}, {Sitarski}, {Ghez}, {Morris},
  {Hees}, {Do}, {Lu}, {Naoz}, {Boehle}, {Martinez}, {Chappell}, {Sch{\"o}del},
  {Meyer}, {Yelda}, {Becklin}, \& {Matthews}}]{Witzel17}
{Witzel}, G., {Sitarski}, B.~N., {Ghez}, A.~M., {et~al.} 2017, \apj, 847, 80

\bibitem[{{Witzel} {et~al.}(2018){Witzel}, {Martinez}, {Hora}, {Willner},
  {Morris}, {Gammie}, {Becklin}, {Ashby}, {Baganoff}, {Carey}, {Do}, {Fazio},
  {Ghez}, {Glaccum}, {Haggard}, {Herrero-Illana}, {Ingalls}, {Narayan}, \&
  {Smith}}]{Witzel18}
{Witzel}, G., {Martinez}, G., {Hora}, J., {et~al.} 2018, \apj, 863, 15

\bibitem[{{Yelda} {et~al.}(2014){Yelda}, {Ghez}, {Lu}, {Do}, {Meyer}, {Morris},
  \& {Matthews}}]{Yelda14}
{Yelda}, S., {Ghez}, A.~M., {Lu}, J.~R., {et~al.} 2014, \apj, 783, 131

\bibitem[{{Yu} \& {Tremaine}(2003)}]{Yu+03}
{Yu}, Q., \& {Tremaine}, S. 2003, \apj, 599, 1129

\end{thebibliography}

\end{document}